1# Using Mobile Phone Data for Electricity Infrastructure Planning

Eduardo Alejandro Martinez-Cesena, Pierluigi Mancarella, Mamadou Ndiaye, and Markus Schläpfer
*Abstract*—Detailed knowledge of the energy needs at relatively high spatial and temporal resolution is crucial for the electricity infrastructure planning of a region. However, such information is typically limited by the scarcity of data on human activities, in particular in developing countries where electrification of rural areas is sought. The analysis of society-wide mobile phone records has recently proven to offer unprecedented insights into the spatio-temporal distribution of people, but this information has never been used to support electrification planning strategies anywhere and for rural areas in developing countries in particular. The aim of this project is the assessment of the contribution of mobile phone data for the development of bottom-up energy demand models, in order to enhance energy planning studies and existing electrification practices. More specifically, this work introduces a framework that combines mobile phone data analysis, socioeconomic and geo-referenced data analysis, and state-of-the-art energy infrastructure engineering techniques to assess the techno-economic feasibility of different centralized and decentralized electrification options for rural areas in a developing country. Specific electrification options considered include extensions of the existing medium voltage (MV) grid, diesel engine-based community-level Microgrids, and individual household-level solar photovoltaic (PV) systems. The framework and relevant methodology are demonstrated throughout the paper using the case of Senegal and the mobile phone data made available for the 'D4D-Senegal' innovation challenge. The results are extremely encouraging and highlight the potential of mobile phone data to support more efficient and economically attractive electrification plans.

## I. Introduction

Detailed knowledge of the energy needs at relatively high spatial and temporal resolution is crucial for the adequate energy infrastructure planning of a country. This is particularly relevant to the electrification of developing regions where new infrastructure needs to be built to foster socio-economic growth. However, such information is typically limited by the scarcity of comprehensive data on human activities. In this respect, during recent years the increasing availability of mobile phone data has proven to provide unprecedented insights into the mobility patterns of people and the distribution of the population in space and time [1]–[3]. Not surprisingly, this type of data has thus been hinted as promising for the design and operation of 'smart' infrastructures [4] and energy systems [5]. However, to the authors' knowledge no quantitative study has so far investigated the potential applicability of mobile phone data for energy infrastructure planning, particularly in developing countries where cellular network data can actually be much more advanced than energy consumption data. For instance, taking Senegal as a typical case of a developing country, during the last decade its mobile phone usage has increased dramatically from 1.7 million subscribers in 2005 to 13.1 million in 2013, thus covering about 95% of the countries 14 million inhabitants [6]. In stark contrast to this upsurge in mobile communication, about half of the total population still has no access to electricity, and the electrification rate in rural areas is even as low as 28% [7]. Therefore, there is a clear potential to use mobile phone data for predicting a region's energy demand and supporting its electrification process. In particular, compared to current approaches for electricity planning in developing countries that use, for instance, satellite imagery, mobile phone data can provide substantially more accurate information on the spatio-temporal activity centers [2], which could be combined with socioeconomic, geo-referenced or climate data for electrification planning purposes in both urban and rural areas.

On these premises, the aim of this work is the assessment of the potential use of mobile phone data to support rural electrification planning in developing countries. Specific objectives include the assessment of *i*) the suitability of mobile phone data as a proxy for current and future electricity needs and whether *ii*) this information can lead to more economical and more efficient electrification options. To that end, we develop a framework that brings together in an innovative way mobile phone data analysis, socio-economic and geo-referenced data analysis, and state-of-the-art energy infrastructure engineering techniques to quantify the techno-economic feasibility of different centralized and decentralized electrification options in developing countries. The electrification options considered here include extensions of the existing Medium Voltage (MV) grid ("centralized" option), development of diesel engine-based community-level Microgrids, and installation of individual dwelling-level solar

E. A. Martinez-Cesena and P. Mancarella are with the University of Manchester, School of Electrical and Electronic Engineering, Electrical Energy and Power Systems Group, Manchester, M13 9PL UK (e-mail: {eduardo.martinezcesena; p.mancarella}@manchester.ac.uk).

M. Ndiaye is with the Ecole supérieure polytechnique de Dakar UCAD, Centre International de Formation et de Recherche en Energie Solaire, 5085 Dakar-Fann, Senegal (e-mail: emamadoulamine.ndiaye@ucad.edu.sn).

M. Schläpfer is with the Santa Fe Institute, Santa Fe, NM 87501 USA (e-mail: schlaepfer@santafe.edu).

photovoltaic (PV) systems ("decentralized" options). The proposed methodology is clearly demonstrated throughout the report by taking the case of Senegal as representative of developing countries.

The report is organized as follows. The next section provides an overview of the different available 'D4D-Senegal' datasets, as well as the electricity context of Senegal, which provides the baseline for this work. Section III presents a high level description of the electrification planning methodology based on mobile phone data proposed in this work. The methodology involves the assessment of *i*) the energy requirements of Senegal, *ii*) the correlation between mobile phone data and electricity needs, *iii*) the population migration towards electrified areas and *iv*) the electrification potential. These steps are further detailed in Sections IV – VII. Section VIII describes possible follow-up studies that could be derived from this work and Section IX concludes.

## II. OVERVIEW OF THE DATASETS

### A. Mobile phone data

The anonymized mobile phone communication data used in this project was collected in Senegal between January 1, 2013 and December 31, 2013. These data were made available by the telecommunications provider Sonatel and the Orange Group within the framework of the D4D–Senegal challenge [8]. In 2013, Sonatel had 7.4 million mobile phone subscribers in Senegal, corresponding to a market share of about 60%. The data are organized into three sets:

- *Dataset 1* contains the hourly voice and text traffic between each pair of mobile phone towers (total call duration, number of calls and total number of text messages). The geographic location of the 1,666 mobile phone towers is depicted in Fig. 1, which also shows the topology of the electricity transmission and distribution networks. Note that a large number of towers lie outside the reach of the power grid; most mobile phone towers without grid access to electricity are in fact powered by diesel generators [9].
- *Dataset 2* contains the fine-grained mobility patterns of about 300,000 randomly sampled and anonymized users during each consecutive period of two weeks. For each time period, a new sample of about 300,000 users was selected and their trajectories recorded at the mobile phone tower level.
- *Dataset 3* contains the coarse-grained trajectories for about 150,000 randomly sampled and anonymized users during the entire year at the spatial level of Senegal's 123 arrondissements (administrative subdivisions). This dataset is not considered in the present study due to its limitations in the spatial resolution.

A more detailed description of the three datasets is provided in [8].

### B. Electricity consumption and infrastructure data

For the purpose of this project, the national electric utility in Senegal – the "Société Nationale d'Éléctricité" (Senelec) –

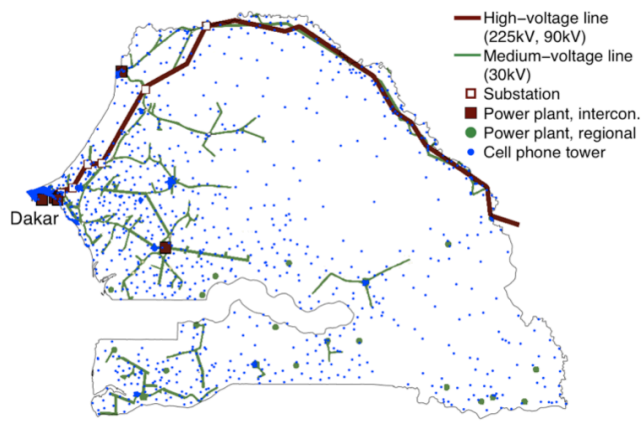

Fig. 1. Existing electricity infrastructure in Senegal and location of the mobile phone towers. The transmission and distribution network as well as the location of the power stations are adopted from [12].

kindly provided us with the hourly electricity consumption data for the entire year of 2013, aggregated at the national level (i.e., 8,760 data points) [10]. The overall yearly electricity consumption was 2,96 TWh. About 80% of the electricity was generated by diesel power plants and the remainder by gas-, steam- and hydro power plants, whereas most of these generators are owned and operated by Senelec [11]. The high-voltage (HV) transmission network consists of 90 kV national and 225 kV supranational lines totaling about 13,000 km in length (see Fig. 1). The 30 kV MV distribution network brings electricity from the transmission network to the consumption centers [13]. Both transmission and distribution networks are again managed by Senelec.

## III. FRAMEWORK AND METHODOLOGY OVERVIEW

As mentioned above, the objective of this work is to build a framework and provide a quantitative assessment methodology for the use of mobile phone data to facilitate rural electrification planning in developing countries in general, and Senegal in particular. Mobile phone use and corresponding mobile phone charging requirements could, in principle, be extrapolated from the mobile phone data. This information could provide key insights into electrification planning of Senegal as mobile phone charging represents, along with lighting, a major energy demand in the country [6], [14]. In addition, mobile phone data could also be used as a proxy for current and future energy needs in a given area and even to estimate the spatio-temporal electricity profiles. This is due to the potential of mobile phone information (particularly if several years' worth of information becomes available) to facilitate the mapping of human activity and migration within the country (e.g., people are more likely to migrate to areas with access to electricity, health and education, thus further increasing energy demands). Both data on human activity and migration can provide an accurate estimation of electricity needs and facilitate more sustainable electrification plans, particularly when combined with other

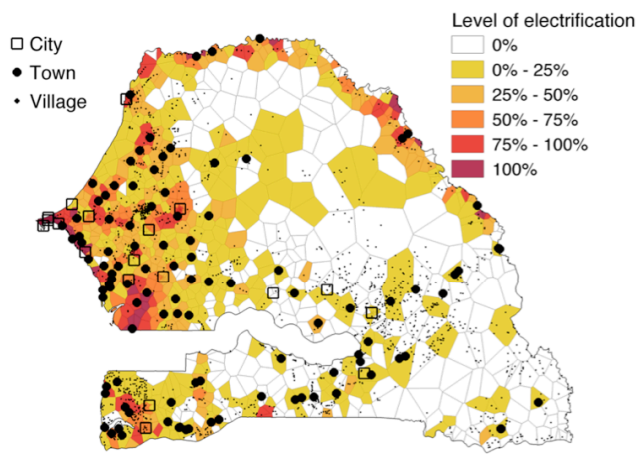

Fig. 2. Level of electrification in the Voronoi polygons defined by the location of the mobile phone towers. The locations of the settlements (cities, towns, villages) are adopted from OpenStreetMap.

TABLE I
EXAMPLE OF THE MAXIMUM AMOUNT OF INSTITUTIONS/SERVICES THAT ARE CONSIDERED FOR AVERAGE VILLAGES OF DIFFERENT POPULATION SIZES IN SENEGAL [15].

|  | Village size (population) | | | |
| --- | --- | --- | --- | --- |
|  | 500 | 1,000 | 5,000 | 10,000 |
| Hospitals | 1 | 1 | 1 | 2 |
| Schools | 1 | 1 | 2 | 3 |
| Markets | 1 | 1 | 3 | 13 |
| Public Lighting points | 3 | 6 | 50 | 99 |

data sources used in state-of-the-art electrification planning practices [15].

In the light of the above, the proposed framework and assessment methodology comprises the following four steps:
1) Assessment of the energy requirements and consumption characteristics of Senegal;
2) Evaluation of the use of mobile phone data as a proxy for current and future electricity needs via correlation analyses;
3) Estimation of potential future migration of population from non-electrified to electrified areas; and
4) Quantification of centralized and decentralized electrification options considering mobile phone data combined with socio-economic and geo-referenced information.

The assessment of energy requirements and consumption characteristics of Senegal is meant to provide context on the expected energy needs of the mobile phone users whose activity is recorded by the different mobile phone towers. This analysis is supported by socio-economic and geo-referenced information extracted from [15] detailing the population density and average distance between households in each area in Senegal, as well as the access to electricity, health, education, markets and so on. This information is used to further classify the mobile phone data compiled from the different mobile phone towers (i.e., *Dataset 1* and *Dataset 2*), allowing the assessment of the correlation between the human activity and the aggregated electricity profile under different socio-economic conditions.

This study is expected to highlight the conditions that make the mobile phone datasets an accurate proxy for current and future electricity needs and profiles. Afterwards, potential migration trends towards electrified areas within the country are assessed based on the fine-grained mobility data (i.e., *Dataset 2*). Again, this information can provide insights into the future energy needs of an area after it is electrified, thus potentially improving electrification decisions. Finally, all this information derived from the mobile phone data is combined with geo-referenced information to build different state-of-the-art options for electrification, namely, MV grid extensions, development of diesel engine-based (community) Microgrids, and development of dwelling-level PV systems (see [15] for an example of the assessment of electrification options for Senegal based only on geo-referenced information). A detailed description of each of the methodological steps and relevant studies is provided in the next sections.

## IV. ENERGY REQUIREMENTS AND CONSUMPTION CHARACTERISTICS OF SENEGAL

The energy requirements and consumption characteristics currently available for Senegal are derived from the countrywide electricity demand profile, the solar radiation and temperatures in different areas, and the size and location of villages and their access to electricity, health, and educational services. The solar radiation and temperature profiles (8,760 hourly data points for 2013) were obtained from the SoDa solar energy services database [16]. A thorough description of the different types of villages in Senegal, their location, and their access to electricity, education and health services were obtained from a previous electrification study in Senegal prepared for the World Bank [15]. Table I lists typical services considered for villages of different sizes.

Together, this information facilitates the differentiation of the mobile phone data based on the context of the area where the mobile phone towers are located. Therefore, we approximated the reception area of each mobile phone tower by a Voronoi tessellation (i.e., the area corresponding to a given tower comprises all points that are closer to that tower than to any other tower) [1]. As a result, the mobile phone data can be classified based on the level of electrification (or access to education, health services etc.) in each Voronoi polygon, as shown in Fig. 2. This classification is critical to identify the conditions where mobile phone data is a good proxy for energy needs, as will be further discussed below.

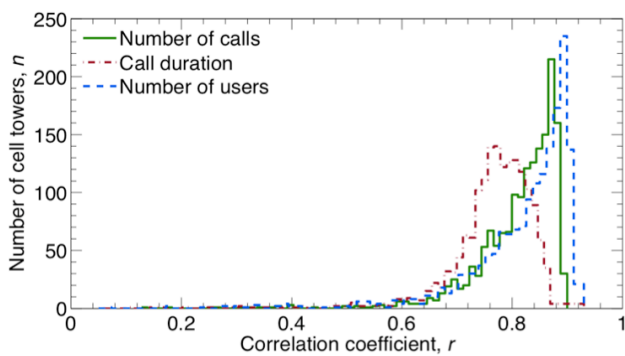

Fig. 3. Histogram of the linear correlation between the electricity load curve and the mobile phone activity within each mobile phone tower area.

## V. CORRELATION STUDIES

In this section, the potential use of mobile phone data, specifically *Dataset 1* and *Dataset 2*, as a proxy for electricity needs is assessed in terms of the correlation between the mobile phone activity at each mobile phone tower and the countrywide aggregated electricity load curve. To that end, we first measured for each mobile phone tower $i$ the linear correlation in terms of the Pearson coefficient $r_i$ with,

$$r_i = \frac{\sum_t \left(A_i(t) - \langle A_i \rangle\right)\left(D(t) - \langle D \rangle\right)}{\sqrt{\sum_t \left(A_i(t) - \langle A_i \rangle\right)^2}\sqrt{\sum_t \left(D(t) - \langle D \rangle\right)^2}} \quad (1)$$

where $A_i(t)$ is the total mobile phone activity during hour $t$ (i.e., number or duration of calls, or number of text messages), $D(t)$ is the countrywide electricity consumption during the same time interval, and $\langle \cdot \rangle$ denotes here the average value over the entire year (8,760 hours). Fig. 3 shows the histogram of the Pearson coefficients, indicating a strong correlation between the hourly mobile phone activity from *Dataset 1* and *Dataset 2*, and the electricity consumption for almost all mobile phone tower areas. The average values over all mobile phone towers for the call duration, number of calls and number of users are $\langle r^d \rangle = 0.76$, $\langle r^n \rangle = 0.8$ and $\langle r^u \rangle = 0.81$, respectively. This result demonstrates that mobile phone data are, in general, a reliable proxy for electricity consumption (and therefore infrastructure needs) in Senegal, even when considering all available data regardless of the characteristics of the villages in the mobile phone tower area (i.e., including non-electrified villages). The correlation study was repeated for areas with different amounts of mobile phone users, population and penetration levels of electricity, health and education. The penetration levels were calculated based on the expected percentage of people living in the mobile phone tower area with access to a given service (e.g., a 50% electrification penetration implies that only half of the individuals living within the mobile phone tower area have access to electricity). This further analysis indicates that low correlations can typically be attributed to the lack of mobile phone data or to areas with either very high or very low electrification levels. More specifically, in 63% of the areas with low correlation (i.e., $r_i < 0.4$), there were only 8 mobile phone users or less. This result is not surprising as mobile phone data cannot be used as a reliable proxy of the electricity needs whenever little or no data are available. Regarding the effects of different electrification levels, electrification was either "too low" (20% or less) or "too high" (80% or more) in 56% of the mobile phone tower areas with low correlation. This result could also be expected considering that the aggregated electricity profile may not be representative for small non-electrified villages, or for large and highly electrified villages. In the latter, for instance, energy-intensive industrial processes that cannot be inferred from human activity profiles may be more widespread.

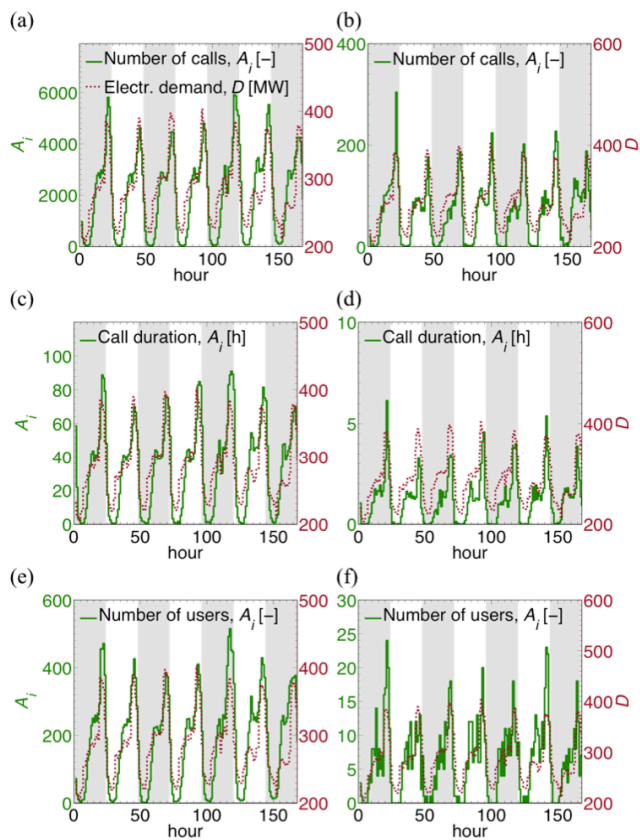

Fig. 4. Comparison of the mobile phone activity (measured as the hourly number of calls, call duration and number of users) with the aggregated electricity load profile (dotted line, in MW) for one week in January 2013 (Monday-Sunday). (a),(c),(e) Mobile phone tower in the city of Dakar. (b),(d),(f) Mobile phone tower in the rural area of Fayil. The shaded areas separate the different days.

Complementary to Fig. 3, Fig. 4 presents examples of the comparison of the mobile phone data and the electricity consumption data in terms of time series profiles for a mobile phone tower in a typical urban area (Dakar) and rural area (Fayil). More specifically, while mobile phone data are actual information from the relevant cellular tower, the electricity profiles are scaled down from the national profile and in proportion to the amount of mobile phone users in the corresponding Voronoi polygon. The visual results confirm the adequacy of mobile phone data as a proxy for electricity needs, which in Senegal seems indeed mostly dictated by human activity (e.g., lighting, mobile phone charging). The figure also shows that good approximations of the electricity profile could be made with either the number of calls, call duration or number of users extracted from *Dataset 1* and *Dataset 2*. Thus, good estimations of the electricity needs could still be made even if the information in the mobile phone datasets were limited.

Overall, the results of the correlation study suggest that, as long as sufficient mobile phone users are available, it is reasonable to use mobile phone data as a proxy for electricity needs under most conditions. Furthermore, this application of mobile phone data seems to be particularly accurate for the average electrified village. This is especially important for electrification planning as, after being electrified, villages are likely to resemble the average electrified village. Accordingly, the results of the correlation study highlight that the use of mobile phone data and a scaled version of the aggregated electricity profile are reasonable for electrification planning.

## VI. MIGRATION STUDIES

In the long-term, the electricity needs of a village are dependent on the expected population growth and migration in the area. Traditionally, population growth and migration have been estimated via census data, which can be enhanced using satellite imagery. Nevertheless, emerging literature suggests that mobile phone data can be used to increase the accuracy of existing population mapping techniques [2].

Several years' worth of mobile phone data beyond *Dataset 2* would be needed to estimate population growth and migration in a given mobile phone tower area with a reasonable level of accuracy. However, considering that the main aim of this work is to illustrate the applicability of mobile phone data to enhance current electrification practices, it is assumed here that the available information in *Dataset 2* suffices for a first estimation of migration to a mobile phone tower area (population growth is taken as 2.3% [15]). Future studies could improve the accuracy of the population mapping (including population growth estimations) should the required information become available.

In order to estimate the potential number of migrants attracted to different villages, the mobility patterns of mobile phone users were calculated based on *Dataset 2*. More precisely, for each mobile phone user we first determined the home location according to [17] and then identified all Voronoi polygons visited throughout the year. Subsequently, we aggregated the number of users that visited a given polygon, providing us with the total number of trips to that area. Finally, we binned the number of trips by the distance of the visitors' home location and normalized it by the total number of trips. We applied the same procedure to determine the number of visits *originating* from a given area. Fig. 5 shows a sample of the results based on electrified and non-electrified areas. Our migration study shows that, as expected, the average amount of travels to an area decreases rapidly as distance increases. More importantly, people seem to travel longer distances to electrified areas than to those without access to electricity (qualitatively similar results apply to access to health and education). The relative difference between the number of people coming from and going to an area, averaged over all possible distances, is taken as the expected migration to an area. Accordingly, migration to electrified areas is assumed here to be between 7% and 13%. This range suggests that the attractiveness of a village is expected to increase if it offers access to electricity.

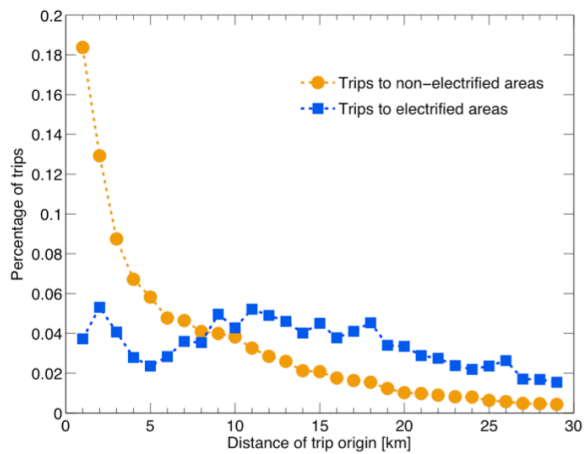

Fig. 5. Normalized histogram of the number of trips to electrified and non-electrified areas, binned by travel distance.

## VII. ELECTRIFICATION STUDIES

In this Section, the energy needs and profiles derived from the mobile phone data, combined with the geo-referenced information extracted from [15], are used for the assessment of three possible electrification options, namely, MV grid extensions, Low Voltage (LV) community-level Microgrids powered by diesel generators, and dwelling-level PV generators. The different options were assessed based on their Net Present Costs (NPC), considering a planning horizon of 10 years and a discount rate of 10%, as recommended by the World Bank [15].

### A. Grid extension

Traditional MV grid extensions involve installing additional MV lines that interconnect the consumption centers with the existing grid, as well as transformers and LV grids to supply the rural villages. This alternative can be particularly attractive to supply large villages near the existing grid, but it may become less economically attractive for smaller villages far from the grid.

The cost of the grid extension (GE) is calculated as the NPC denoted by (2), (3) and (4). The inputs for the relations are the specific characteristics of the village to be electrified (i.e., electricity consumption and peak) taken from the scaled electricity profile (adjusted for migration), the geo-referenced information extracted from [15] and the parameters given in Table II.

TABLE II
ECONOMIC AND TECHNICAL PARAMETERS FOR THE MV GRID EXTENSION
ELECTRIFICATION ALTERNATIVE [15]

| | |
|---|---|
| *MV line* | |
| Cost | 8M CFA/km |
| O&M cost | 2% (of the total investment) |
| *Transformer* | |
| Capacity | 5 kVA – 100 kVA |
| Cost | 2M CFA – 4M CFA |
| O&M cost | 3% |
| *LV line* | |
| Costs | 6M CFA/km |
| O&M costs | 3% |
| *Generation* | |
| Generation costs | 83.4 CFA/kWh |
| Losses | 15% |

TABLE III
ECONOMIC AND TECHNICAL PARAMETERS FOR THE DIESEL ENGINE-BASED
MICROGRID ALTERNATIVE [15]

| | |
|---|---|
| *Generation* | |
| Capacity | 10kVA – 50kVA |
| Cost | 6.4M CFA – 9.8M CFA |
| Generation costs | 216CFA/kWh |
| O&M cost | 5% |
| Life time | 5 years |
| *LV line* | |
| Costs | 6M CFA/km |
| O&M costs | 3% |
| Losses | 5% |

$$NPC_{GE} = I_{GE} + \sum_{i=0}^{T} \frac{C_{GE}(D_i + L_{MV,i}) + M_{GE}}{(1+d)^i} \quad (2)$$

$$M_{GE} = M_{MV} + M_{TR} + M_{LV} \quad (3)$$

$$I_{GE} = I_{MV} + I_{TR} + I_{LV} \quad (4)$$

where $I_{GE}$ is the total investment cost in CFA[1], $C_{GE}$ is the electricity generation cost in CFA/kWh, $D_i$ is the annual demand of the village in kWh/year, $L_{MV,i}$ represents annual power losses in kWh/year, $d$ is the discount rate, $i$ denotes a year, $T$ represents the planning horizon (years), $M_{GE}$, $M_{MV}$, $M_{TR}$, and $M_{LV}$ denote the annual operation and maintenance (O&M) costs in CFA/year associated with the total investment cost, as well as with the MV line, transformer, and LV line, respectively. The parameters $I_{MV}$, $I_{TR}$ and $I_{LV}$ denote the investment costs in CFA associated with the MV line, transformer, and LV lines, respectively. It is important to note that the investment costs are a function of the length of cable and capacity of the transformer to be installed.

The distance between the existing MV network and the villages was estimated based on the geo-referenced information from [15] and using an iterative procedure to find the minimum length between villages and existing network connection points. The length of the LV network was calculated by assuming that the mean distance between households varies between 8 m in villages with more than 5000 individuals to 30 m for villages with less than 500 people [15]. This distance is assumed to increase by up to 50% for dwellings located far from the center of the village.

*B. Diesel engine-based Microgrid*

Instead of extending the MV grid to the location of a village, it is possible to install a group of distributed generation units (diesel generators in this case) and a LV Microgrid to supply a village located far away from the existing grid. The NPC of the Microgrid (MG) electrification alternative ($NPC_{MG}$) is calculated with (5), (6), (7) and (8). Similarly to the previous case, the inputs for the equations come from the characteristics of the village to be electrified (i.e., electricity consumption and peak), the geo-referenced information from [15] and Table III.

$$NPC_{MG} = I_{MG} - S_G + \sum_{i=0}^{T} \frac{C_{MG}(D_i + L_{MG,i}) + M_{MG}}{(1+d)^i} \quad (5)$$

$$M_{MG} = M_G + M_{LV} \quad (6)$$

$$I_{MG} = I_{LV} + \sum_{i=1}^{floor\left(\frac{T}{LT_G}\right)} \frac{I_G}{(1+d)^{i \cdot LT_G}} \quad (7)$$

$$S_G = \frac{I_G}{(1+d)^{floor(T/LT_G) \cdot LT_G}} \cdot \frac{T - floor(T/LT_G)}{LT_G} \quad (8)$$

where $I_{MG}$ is the total investment cost (in CFA), $S_G$ is the salvage value of the generator, $C_{MG}$ is the generation cost in CFA/kWh, $L_{MG,i}$ represents the annual power losses (in CFA/year) and $LT_G$ represents the lifetime of the generator (in years). The parameters $M_{MG}$ and $M_G$ denote the annual O&M costs in CFA/year associated with the total investment and generator, respectively, and $I_G$ represents investments in generators. It is important to highlight that several investments in generators may be needed throughout the planning horizon.

*C. PV systems*

Off-grid PV systems supplying individual households tend to be less economically attractive than the other technologies under normal conditions. However, due to their modularity, independent PV systems can be installed in each household without the need of an LV network. Thus, this option can become economically attractive for low population villages where dwellings are located far apart.

The NPC of a PV system ($NPC_{PV}$) is calculated with (9), (10), (11) and (12). The inputs for these equations were taken from the characteristics of the villages to be electrified, the geo-referenced information from [15] and Table IV.

---

[1] 1 CFA = 0.0012 GBP = 0.002 US$

TABLE IV
OVERVIEW OF THE ECONOMIC AND TECHNICAL PARAMETERS
CONSIDERED FOR A PV SYSTEM ELECTRIFICATION ALTERNATIVE [15].

| *PV Module* | |
|---|---|
| Capacity | 20W – 150W |
| Cost | 88k CFA – 660k CFA |
| O&M costs | 1% |
| *Batteries* | |
| Capacity | 14Ah – 38Ah |
| Cost | 40k CFA – 70kCFA |
| O&M cost | 1% |
| *Converter* | |
| Costs | 28k CFA |
| O&M costs | 1% |

$$NPC_{PV} = I_{PV} - S_B + \sum_{i=0}^{T} \frac{M_{PV}}{(1+d)^i} \tag{9}$$

$$M_{PV} = M_{MO} + M_{CV} + M_B \tag{10}$$

$$I_{PV} = I_{MO} + I_{CV} + \sum_{i=0}^{floor\left(\frac{T}{LT_B}\right)} \frac{I_B}{(1+d)^{i \cdot LT_B}} \tag{11}$$

$$S_B = \frac{I_B}{(1+d)^{floor(T/LT_B) \cdot LT_B}} \cdot \frac{T - floor(T/LT_B)}{LT_B} \tag{12}$$

where $I_{PV}$ is the total investment cost in CFA, $S_B$ is the salvage value of the batteries and $LT_B$ represents the lifetime of the batteries (in years). The parameters $M_{PV}$, $M_{MO}$, $M_{CV}$ and $M_B$ denote the annual O&M costs in CFA/year associated with the total investment, modules, converter and batteries, respectively, and $I_{MO}$, $I_{CV}$ and $I_B$ represent investments in modules, converters and batteries, respectively.

We applied the detailed simulation technique proposed in [18], based on an estimated electricity profile (i.e., a scaled version of the aggregated electricity profile) and the technical characteristics of the converters and batteries, in order to optimize the design (i.e., type and amount of modules and batteries) of the PV system for each residential dwelling and institutional building such as a hospital, school, and so forth. The simulation approach allows the assessment of the lifetime of the batteries, which facilitates the identification of low cost PV designs. This is because it may be convenient to oversize the PV module or battery array if it leads to an increased lifetime for the batteries. The lifetime of the batteries is estimated with the Ah throughput model denoted by (13), (14) and (15), see [18] for a detailed description.

$$LT_B = \frac{AhT}{\sum_{t=1}^{8760} i_{eff,t}} \tag{13}$$

$$AhT = DOD_{rated} \cdot CAP_{rated} \cdot N_{rated} \tag{14}$$

$$i_{eff,t} = \frac{i_{rated}}{CAP(i_t)} \cdot \frac{N_{rated}}{N(DOD_t)} \cdot i_t \tag{15}$$

where $AhT$ is the expected throughput of a battery (in A), $i_t$ is the discharge current (in A), and $i_{eff,t}$ is the effective discharge current (in A). The parameters $DOD_{rated}$ and $DOD_t$ (in percent) are the depth of discharge under rated conditions and at a given time, respectively, $CAP_{rated}$ and $CAP(i_t)$ (in A) are the capacity of the battery subject to rated conditions and a given discharge current, respectively, $N_{rated}$ and $N(DOD_t)$ are the amount of operational cycles subject to rated conditions and a given depth of discharge, respectively, and the subscript $t$ denotes a given hourly time step within a year.

### D. Electrification option results and discussion

The three different options and relevant technologies discussed above were assessed for the electrification of the villages in Senegal, classified based on their population and access to services such as education, health and markets as extracted from [15]. Each electrification option was assessed based on parametric scenarios and taking the cheapest alternative as the recommended technology. In addition, the option to install small PV systems in small villages to supply only demand from mobile phones charging and lighting (which are the current main energy needs, as discussed in [6]) was considered too.

The parametric scenarios were formulated assuming different levels of electrification (from 20% to 80%) and population growth due to migration (from 0% to 13%), as suggested by our migration studies. The electrification levels represent potential electrification targets, such as the current target in Senegal to achieve 60% electrification of rural areas [15]. It is considered that electrification levels for each newly electrified village will be the same. For instance, if a 50% electrification level is considered, only 50% of the households in every village will be electrified, which would correspond to the dwellings in the densest area of each village and nearest to the center of the settlement. The results highlight that each of the electrification options can outperform the others under a specific set of conditions, as shown in Fig. 6.

Fig. 6(a) shows the costs per household associated with individual PV and diesel engine-based Microgrids, subject to different village population sizes and electrification levels. The PV systems are only deemed more attractive than the Microgrid for small villages where houses may be dispersed and the installation of a LV network would be too expensive, and for low electrification levels where the installation of a diesel engine may not be justifiable. The use of small PV systems is the only economically viable option for the electrification of small villages where mobile phone charging and lighting may be the main electrical load. This result is consistent with existing literature [15]. It is important to note that the results regarding the PV system present non-monotonic, but well defined increasing trends due to the nonlinearity and integer nature of the arrays of PV modules and batteries (i.e., it is not possible to install a fraction of a PV module or battery).

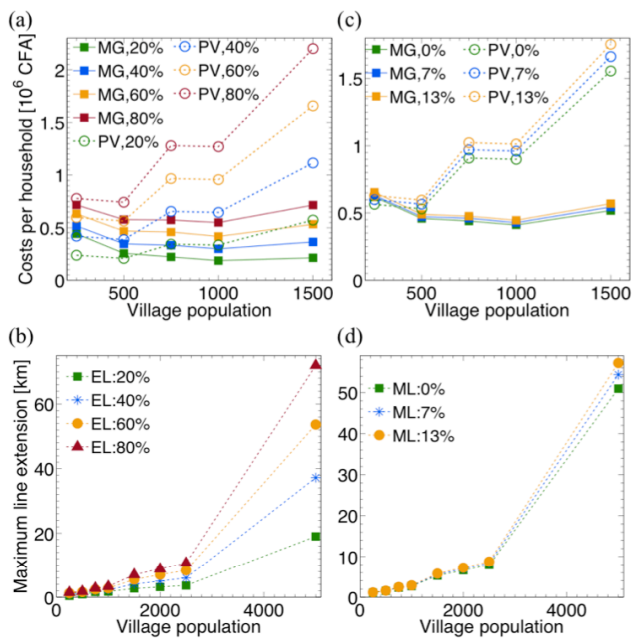

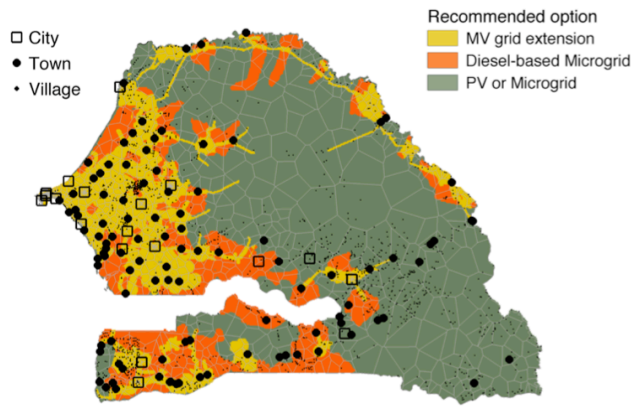

Fig. 6. (a) Costs of PV systems (PV) and diesel-based Microgrids (MG) for different electrification levels. The curves are averages over three migration levels (0%,7%,13%). (b) Maximum MV grid extensions that are economically competitive with PV and MG for different electrification levels. The curves are averages over different migration levels as in (a). (c) Costs of PV and MG for different migration levels. The curves are averages over different electrification levels (0%,10%,...,90%,100%). (d) Maximum economically competitive MV grid extensions for different migration levels, averaged as in (c).

Fig. 7. Electrification recommendations considering 60% electrification levels and 13% migration to electrified villages.

Fig. 6(b) shows the longest possible MV grid extension that would still be economically competitive with other alternatives (i.e., PV systems and Microgrids). This is calculated by first estimating the costs corresponding to a Microgrid and a PV system in the specific village, and then estimating the maximum potential MV grid extension (using (2) to (4)) that would be cheaper than the other technologies. The results suggest that it is more attractive to extend the grid when this enables high electrification levels for large villages, while it is attractive to use this option to electrify small villages when they are located in the proximity of the existing grid. It is important to note that investments in *upstream* generation and grid upgrades that might be needed to support MV line extensions are not considered in this study. Such costs would make line extensions – particularly when planning many line extension or long line extensions – for big villages less attractive, thus a decrease in the maximum recommended line extension (particularly after about 20 km) would be expected in the results.

Fig. 6(c) and Fig. 6(d) show the effects that migration could have on the preferred electrification option. It can be observed that the effects of migration on electrification planning are modest compared with those associated with different electrification levels. However, migration can still play a big role on the identification of small villages where PV systems are deemed the only feasible alternative. This is due to the potential of migration to increase (or decrease) the population of a village. It is worth noting that these results only highlight the potential impacts of migration in a parametric way and that additional research would be needed to better quantify the actual migration potential in Senegal.

*E. Recommendations for the electrification of Senegal*

Finally, a graphical representation of the results presented in Fig. 6 and applied to all the villages in the geo-referenced dataset extracted from [15] is presented in Fig. 7. As discussed above, for each relevant option the length of required MV extension was calculated based on the distance between the existing grid and the village, whereas the length of the LV networks were estimated based on the distance between households in each village, electrification level and growth due to migration. The electrification levels and population growth in combination with peak demand and total energy consumption, estimated from the mobile phone data, were used when sizing the PV arrays, generators and transformers.

Fig. 7 was calculated assuming 60% electrification and 13% migration and shows the zones where it would be more economically attractive either to extend the MV grid, or to install a diesel engine-based Microgrid or individual PV systems. In the green (dark) zones, there are two potential technologies to be deployed, namely Microgrids and PV systems. The latter are recommended for the smallest villages or for areas where only lighting and mobile charging infrastructure is to be electrified. Note that similar figures could be readily produced for alternative electrification and migration scenarios.

VIII. OPENINGS TO FURTHER WORK

If more detailed electricity and mobile phone data were available for longer observation periods, further work could be done to improve the analysis carried out here. For instance, a more detailed assessment of the upstream costs in the case of the MV extension option could be performed; also, a more in-depth analysis of possible changes in the mobile phone activity profiles due to the electrification of a settlement could be carried out, so as to improve the energy planning option assessment; finally, more electrification options could be considered, such as those based on wind turbines and/or fuel cells.

Further, given the novelty of the topic, there is significant research that could still be carried out for electrification planning and for other infrastructure-related applications based on mobile phone data. For instance, additional detailed mobile phone datasets covering several years would provide better proxies for electricity needs and population migration, particularly when combined with corresponding electricity consumption profiles for different areas. Moreover, the hourly activity curves derived from mobile phone data could be compared with environmental data such as hourly solar radiation or wind speeds [19] to quantify in more detail the potential for PV or wind power in a given area and estimate the need for energy storage.

More advanced studies could also be carried out in the context of developed countries. For instance, the dynamic population mapping derived from mobile phone data could be used for assessing the number of people that would be affected by a potential power blackout. Such 'risk-maps' could inform the extension and operation of existing power grids. Finally, as an example for future infrastructure scenarios, the derived people flows could provide valuable information on where to place charging stations for plug-in electric vehicles.

## IX. CONCLUSION

This work has introduced an original framework and relevant assessment methodology to use mobile phone data for the enhancement of electrification practices in developing countries. This framework brings together in an innovative way mobile phone data analysis, socio-economic and geo-referenced data analysis, and state-of-the-art energy infrastructure engineering techniques. More specifically, mobile phone data have been used as a proxy for current and future electricity requirements in different areas. Subsequently, this information was used to quantify the techno-economic feasibility of different centralized and decentralized electrification options in Senegal.

The study shows that mobile phone data can be an accurate means to estimate the energy consumption, peak demand and even the electricity profile of different regions. This information, in turn, has proven to be able to facilitate detailed technical and economic assessments of the considered electrification options, namely, MV grid extension, diesel engine-based Microgrids, and individual household PV systems. The results clearly demonstrate how our framework and methodology can be adopted to quantify how the use of mobile phone data can effectively support electrification plans in developing countries with scarce information on local energy consumption and limited electrification in many areas. Several possible future extensions of the current work have also been discussed in detail, predicated on more extensive energy and mobile phone data.

ACKNOWLEDGMENT

The authors acknowledge Papa Alioune Ndiaye (Ecole supérieure polytechnique de Dakar) for providing the country-wide electricity consumption data for 2013. M. Schläpfer thanks Luis Bettencourt, Paul Hines and Seth Blumsack, and acknowledges the Minerva Research Initiative.